\documentclass[preprint2]{aastex}
\usepackage{graphicx}
\usepackage{longtable,lscape}
 
\shorttitle{Stellar parameters of K and M dwarfs}
\shortauthors{Houdebine et al.}

\begin{document}


\title{The Mass-Activity relationships in M and K dwarfs. I. 
Stellar parameters of our sample of M and K dwarfs
\thanks{Based on Gaia DR2 and Hipparcos parallax 
measurements.}}

\author{\'Eric R. Houdebine\altaffilmark{1,2}, D.J. Mullan\altaffilmark{3}, J.G. 
Doyle\altaffilmark{1}, Geoffroy de La Vieuville\altaffilmark{2}, C.J. 
Butler\altaffilmark{1}, F. Paletou\altaffilmark{2}}

\altaffiltext{1}{Armagh Observatory, College Hill, BT61 9DG Armagh, 
Northern Ireland}

\altaffiltext{2}{Universit\'e Paul Sabatier, Observatoire Midi-Pyr\'en\'ees, 
Cnrs, Cnes, IRAP, F--31400 Toulouse, France}

\altaffiltext{3}{Department of Physics and Astronomy, University of Delaware, 
Newark, DE 19716, USA}













\begin{abstract}
Empirical correlations between stellar parameters such as rotation or radius 
and magnetic activity diagnostics require estimates of the effective 
temperatures and the stellar radii. The aim of this study is to propose simple 
methods that can be applied to large samples of stars in order to derive 
estimates of the stellar parameters.

Good empirical correlations between Red/Infra-Red colours (e.g. (R-I)$_C$) and 
effective temperatures have been well established for a long time (e.g. Veeder 
1974, Bessell 1979, Leggett 1992). 
The more recent (R-I)$_C$ colour-$T_{eff}$ correlation using the data 
of Mann et al. (2015: hereafter M15) and Boyajian et al. (2012: hereafter B12) 
shows that this colour can be applied as a temperature estimate for large 
samples of stars. We find that the mean scatter in $T_{eff}$ relative to the 
(R-I)$_C$-$T_{eff}$ relationship of B12 and M15 data is only $\pm 3\sigma 
=$44.6~K for K dwarfs and $\pm 3\sigma =$39.4~K for M dwarfs. These figures 
are small and show that the (R-I)$_C$ colour can be used as a 
first guess effective temperature estimator for K and M dwarfs.

We derive effective temperatures for about 1910 K and M dwarfs using the 
calibration of (R-I)$_C$ colour-$T_{eff}$ from B12 and M15 data. We also 
compiled $T_{eff}$ and metallicity measurements available in the literature 
using the VizieR database.  

We determine $T_{eff}$ for 441 stars with previously unknown effective 
temperatures. We also identified 21 new spectroscopic binaries and 1 triple 
system from our high resolution spectra.

\end{abstract}
\normalsize


\section{Introduction}

The determination of fundamental stellar parameters ($T_{eff}$, $R_{*}$, and 
[M/H]) are essential to many astrophysical studies such as the determinations 
of exoplanet properties, the comparison to stellar interior models, the 
determinations of Rotation-Activity Correlations (RACs) and another type of 
magnetic activity related correlations presented in this series of papers, 
e.g., the Mass-Activity empirical Correlations (MACs thereafter). In 
exoplanet studies, the precise determinations of the stellar parameters are 
required to establish the radius of a transiting planet. Also, the mass of an 
exoplanet detected with a given orbital period scales as the mass of the star 
to the power 2/3. 

The determinations of estimates of stellar radii are also important in order 
to obtain RACs and MACs with a minimum amount of scatter (e.g. Houdebine et al.
 2017, Houdebine et al. 2019, in preparation). We estimate that the 
uncertainties in the determination of rotation periods are due to the 
uncertainties in stellar radii determinations by an amount of about 50\% .
In the present series of papers, estimates of stellar radii are also central 
in order to obtain meaningful empirical correlations between magnetic activity 
diagnostics, i.e. the (Ca\,{\sc ii} resonance lines and $H_{\alpha}$ surface 
fluxes or $R'_{HK}$), and various quantities related to the stellar radius: 
$L_{Bol}$, $M_V$, $M_{*}$ or $R_{*}$. In this paper, we focus on quantities 
which are relevant to determining values of stellar radii and $T_{eff}$s, for 
a large sample of 1910 K and M dwarfs from dK3 to dM7. We shall postpone the 
discussion of activity indices to Paper II, when the MACs will be constructed.

Stellar parameters of K and M dwarfs were sometimes estimated by direct 
comparison to models (e.g. Paletou et al. 2015, Casagrande et al. 2008). 
However, in the case of M dwarfs there are still numerous atomic and 
molecular lines that are not included in the currently existing models, 
although there is a continuing effort to improve the models (e.g. Allard 
et al. 2015). Also the current models do not include possible NLTE effects 
nor the presence of spots and the chromosphere. The stellar 
radii obtained from the studies of low-mass eclipsing binaries (LMEBs) 
indicate that stellar models tend to under-predict the radii (e.g., Kraus et 
al. 2014, Spada et al. 2013). Mullan \& McDonald (2001) showed that these 
discrepancies may be due to the high activity levels of the LMEBs. On the 
contrary, M15 found some consistency between their derived values of 
$L_{Bol}$, $R_{*}$, $M_{*}$ and $T_{eff}$ and those predicted by their model 
calculations. Specifically, M15 found that their models systematically 
over-predict $T_{eff}$ and under-predict $R_{*}$ by -2.2\% and 4.7\% 
respectively for their hotter stars, in agreement with previous comparisons of 
models and LMEBs and single main-sequence field stars. M15 also found that 
for values of $T_{eff}$ below 3500~K, where stars on the main sequence are 
predicted to be fully convective, $F_{Bol}$ was systematically underestimated 
by the models at the 0.2\% level. For stars hotter than 3500~K, the models 
overestimate $F_{Bol}$ by about 0.4\%. M15 also reported systematic 
differences between their model-based masses and those from Delfosse et al. 
(2000). The models predict systematically lower masses above 0.50~$M_{\odot}$ 
and systematically higher masses below this threshold. They propose that their 
model-derived masses are more precise than those from the Delfosse et al. 
(2000) relation.

B12 compared the luminosity-temperature, luminosity-radius, temperature-radius 
and mass-radius empirical relationships for their sample of 33 K and M dwarfs 
to the models of Padova, Dartmouth, BCAH and Yonsei-Yale  (Girardi et al. 
2000, Dotter et al. 2008, Baraffe et al. 1998, Demarque et al. 2004). They 
found that the Dartmouth and BCAH models reproduce the trends of their data 
the best. In their temperature-radius plane they found that there is
 a lot of scatter in the radius of a star for a given temperature. B12 also 
found that models over-predict temperatures by an average of 6\%. For 
radii less than 0.7~$R_{\odot}$ they found that models under-predict radii by 
about 10\%, and the difference between their observations and models increases 
with decreasing radii up to $\sim$50\% for radii of $\sim$0.4~$R_{\odot}$. 

Houdebine (2008, Paper VII) found a relatively tight correlation between [M/H] 
and $R_{*}$ in a sample of M2 dwarfs. [M/H] decreases with decreasing  $R_{*}$ 
in normal dwarfs and subdwarfs, such that all subdwarfs have very low [M/H] 
and $R_{*}$. In Houdebine et al. (2016a) we performed extensive compilations 
of [M/H] from the literature for late-K, M3 and M4 dwarfs and correlated [M/H] 
and $R_{*}$. These empirical radius-[M/H] correlations emphasize that stellar 
radii diminish markedly with decreasing [M/H]. These empirical correlations 
can explain (at least in part) the scatter observed in the temperature-radius 
plane in B12 and M15.

As far as the current series of papers is concerned, we aim in Houdebine et 
al. (2019, Paper II, in preparation) to constrain the efficiency of the 
dynamo mechanisms as a function of stellar radius and mass. An initial 
empirical correlation was found between the Ca\,{\sc ii} line mean EW and 
absolute magnitude $M_V$ by Houdebine (1996) for a sample of M2 dwarfs. Later, 
Houdebine\& Stempels (1997, Paper VI) also found correlations between the 
Ca\,{\sc ii} line mean EW and $M_V$ for a larger sample of M2 dwarfs. In an 
effort to constrain the dynamo mechanisms at spectral type M2, Houdebine (2011,
 Paper XV) also found correlations between the Ca\,{\sc ii} line mean EW, $M_V$
 and [M/H] for a much larger sample of M2 dwarfs. He also found a correlation 
between the Ca\,{\sc ii} line surface fluxes corrected for metallicity 
effects and radius: $F_{CaII}\propto R_{*}^{3.6}$. A consequence of this is 
that $L_{CaII}$ grows roughly as the power of 5.6 of the stellar radius in 
their M2 dwarf sample. Similar correlations were found with the $H_{\alpha}$ 
line diagnostic.

In the current series of papers, preliminary results indicate that the 
Ca\,{\sc ii} line luminosity, $L_{CaII}$, for the samples of low activity stars
 from Houdebine et al. (2017) increases roughly as the power of 8 of the 
stellar radius, a value significantly larger than that found in Paper XV. For 
the active stars, $L_{CaII}$ increases as $R_{*}$ to the power of 6.6. This 
means that the Ca\,{\sc ii} luminosity depends most sensitively on the stellar 
radius (or stellar mass). The power law index for $R_{*}$ is much larger (in 
magnitude) than that of the effect of the corresponding index for the 
dependence of $L_{CaII}$ on the rotation period. For the latter, the absolute 
value of the power law index is largest for dM4 stars (where the magnitude is 
3.77, see Houdebine et al. 2017) and smallest for dK5 stars (for which the 
magnitude is 0.70). These results suggest that the efficiency of the dynamo 
mechanism may depend primarily on the stellar radius (or mass) in M dwarfs.

We stress that the contents of the present paper are only a first step in a 
two-step process which aims to derive MACs for our complete stellar sample. 
The second step in this process will be discussed in Paper II. The present 
paper focusses on obtaining estimates of the data for the ``Radius" or 
"Mass" axis of the MACs. The subsequent paper will focus at first on obtaining 
reliable data for the ``Activity" axis of the MACs. Once reliable data are 
available for both axes, a search will then be undertaken (in Paper II) to 
determine the  correlation between Radius or Mass-related parameters and 
Activity-related diagnostics for our stellar sample.

In the aim to gather estimates of the activity parameter for a large sample of 
M and K dwarfs, we are currently gathering high resolution/high S/N spectra 
with the SOPHIE (Haute-Provence Observatory) and NARVAL (Pic-du-Midi 
Observatory) high resolution spectrographs. We aim to gather about 500 such 
spectra of stars that have never been observed in high resolution spectroscopy 
so far. We presently have obtained high resolution spectra for about 350 stars.

The present paper focuses on the determination of effective temperature
 based mainly on a correlation between the (R-I)c color and $T_{eff}$. 
To use this correlation we have relied on the measurements of (R-I)c colors 
reported by B12 and M15. We believe that B12 provide probably the most 
accurate determinations of $T_{eff}$ for a sample of 33 K and M dwarfs: 
our belief is based on the fact that the method used by B12 is nearly 
independent of stellar models. As is widely recognised, models are 
inevitably subject to uncertainties based on incomplete opacity sources 
in the calculations. Moreover, the models also assume that LTE conditions 
hold true throughout the atmosphere. In contrast to B12, the approach 
used in M15 is to rely to some extent on deriving stellar parameters 
by taking advantage of certain models. However, the authors of M15 do 
not rely solely on the models: they also use the constraints which are 
provided by precision interferometry in order to derive the stellar 
parameters. Therefore, in the present paper, by relying in part on M15, 
it is important to note that interferometric measurements of stellar radii 
provide a first step in the process by means of which we derive stellar 
parameters. In a second step, we compile $T_{eff}$ and [M/H] measurements 
from the literature for most of our targets. When we compare our $T_{eff}$ 
with the values which already appeared in the literature, we find that  
the agreement between our results and those in the literature is 
generally good. As a third step in our process, we also use the GAIA 
DR2 parallaxes in order to ensure more reliable estimates of the stellar 
radii.

\begin{figure*} 
\vspace{-1.5cm}
\begin{centering}
\hspace{-2.8cm}
\includegraphics[width=14cm,angle=-90]{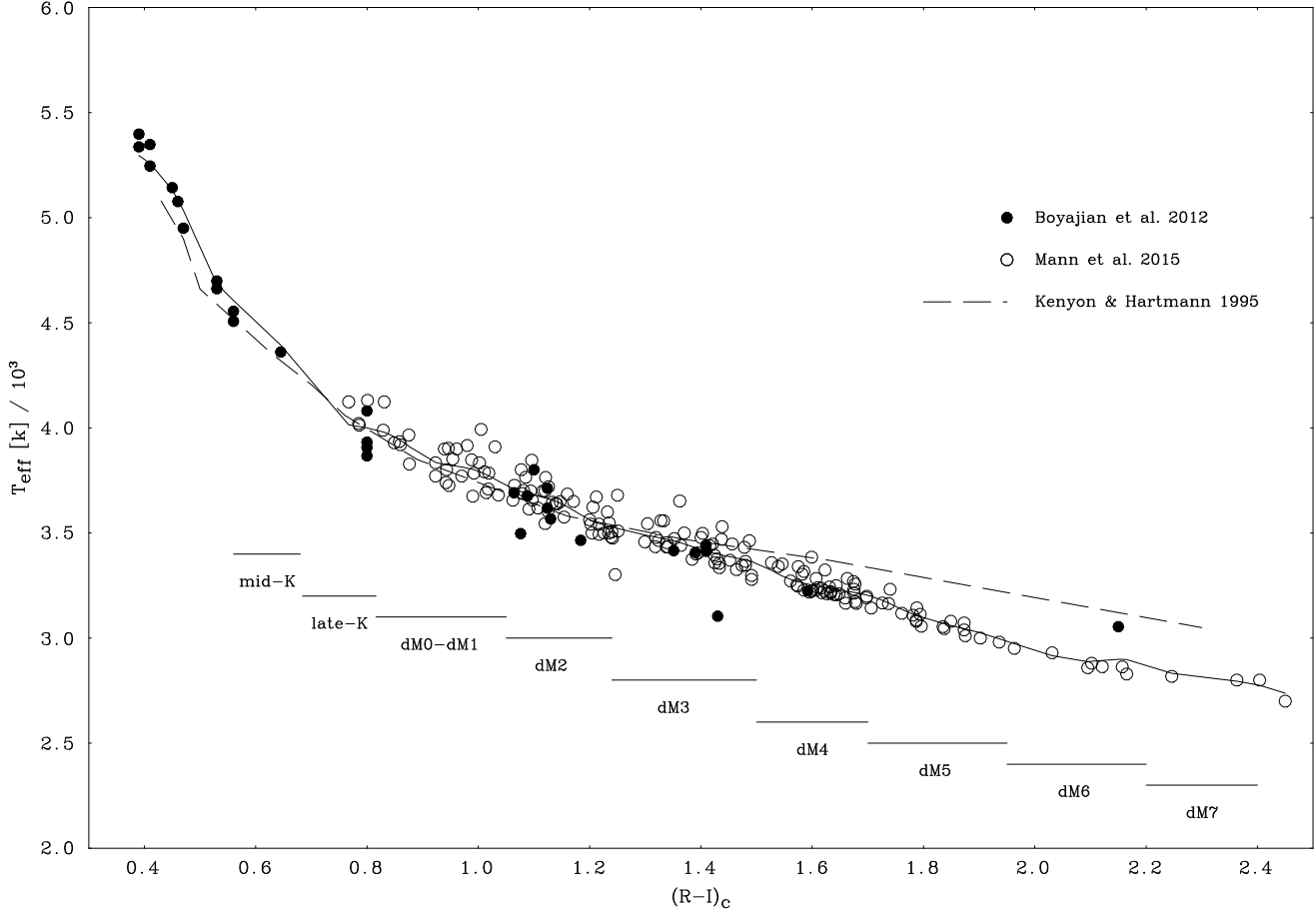}
\vspace{-0.5cm}
\end{centering}
\caption[]{Values of $T_{eff}$ as a function of the (R-I)c color for the data 
of B12 (filled circles) and M15 (open circles) and a smoothing of these data 
(dotted line). Also shown for comparison: the calibration of Kenyon and 
Hartmann (1995). This later calibration agrees well with the more recent 
measures of Mann et al. (2015) up to (R-I)c=1.3, but tends to overestimate 
$T_{eff}$ for later spectral types. We overplot the (R-I)c domains of our 
samples of stars from mid-K to dM7.}
\end{figure*}

\section{Selection of the samples of stars}

It is now well established that the red/infra-red colours are well 
correlated with effective temperatures in late-K and M dwarfs (e.g. Veeder 
1974, Mould \& Hyland 1976, Bessell 1979, Leggett 1992, Ramirez \& Melendez 
2005, Boyajian et al. 2012, Mann et al. 2015, Houdebine et al. 2017). Based 
on our previous studies, we have found that the most suitable initial selection
 parameter when we wish to identify a homogeneous sample of K or M dwarfs 
belonging to a specific sub-type is the (R-I) color: this color is sensitive 
to $T_{eff}$, but less so to metallicity (e.g. Leggett 1992, Ramirez \& 
Melendez 2005). Moreover, broad-band colours of high precision (typically 
of the order of 3\% , Leggett 1992) are widely available in the literature 
for many of the cool dwarfs which are of interest to us.

Observations of (R-I)$_{C}$ (Cousin's photometric system) or (R-I)$_{K}$ (Kron 
photometric system) for our samples of K and M dwarfs were taken from the 
following papers: Eggen (1971), Veeder (1974), Eggen (1974), Rodgers \& Eggen 
(1974), Eggen (1976a, 1976b), Mould \& Hyland (1976), Eggen (1978), Eggen 
(1979), Eggen (1980), Weis \& Upgren (1982), Upgren \& Lu (1986), Eggen (1987),
 Booth et al. (1988), Leggett \& Hawkins (1988), Dawson \& Forbes (1989), 
Laing (1989), Bessel (1990), Weis (1991a 1991b), Dawson \& Forbes (1992), 
Leggett (1992), Ryan (1992), Ruiz \& Anguita (1993), Weis (1993), Weis (1996), 
Reid et al. (2004), Koen et al. (2010). 

We selected 8 samples of K and M dwarfs according to their (R-I)$_{C}$ colours:
mid-K sample ((R-I)$_{C}\in$[0.560:0.680], (R-I)$_{K}\in$[0.400:0.500], 
465 stars), late-K sample ((R-I)$_{C}\in$[0.684:0.816], (R-I)$_{K} 
\in$[0.503:0.613], 419 stars), M2 sample ((R-I)$_{C}\in$[1.050:1.240], 
(R-I)$_{K}\in$[0.823:0.972], 520 stars), M3 sample ((R-I)$_{C} 
\in$[1.240:1.500], (R-I)$_{K}\in$[0.972:1.180], 767 stars), M4 sample 
((R-I)$_{C}\in$[1.500:1.700], (R-I)$_{K}\in$[1.180:1.349], 395 stars), M5 
sample ((R-I)$_{C}\in$[1.70:1.95], (R-I)$_{K}\in$[1.349:1.579], 155 stars),
M6 sample ((R-I)$_{C}\in$[1.95:2.20], (R-I)$_{K}\in$[1.579:1.860], 30 stars),
M7 sample ((R-I)$_{C}\in$[2.20:2.40], (R-I)$_{K}\in$[1.860:2.216], 14 stars). 
This represents a total sample of 2765 K and M dwarfs. Most stars in this 
sample are nearby or large proper motion stars. We report to Leggett (1992) 
for a detailed desciption of the Cousin's and Kron photometric system as well 
as for the transformation formulae between the two systems. The error of the 
transformation from the Kron system and the Cousin's system should be less 
than 1\%  (Leggett 1992). We also completed 
these samples of stars with a 9$^{th}$ sample of stars: the M0-M1 sample, 
which includes some stars from the samples of stars from B12 and M15 (see 
Table~1), as well as several stars from initially the M2 sample which were 
found to have higher temperatures (we included in the M0-M1 sample, stars down 
to the spectral type dM1.5).

We note, however, that spectral classification may differ from one author to 
another. In this regard, in previous Papers on dM1 stars (Houdebine 2008, 
Paper VII, Houdebine 2009, Paper XII, Houdebine et al. 2009 Paper XIII, 
Houdebine 2010b Paper IX, Houdebine 2010a Paper XIV, Houdebine 2010c Paper X, 
Paper XV, Houdebine et al. 2012 Paper XIX), we used a different calibration. 
According to their infrared colours and the classification of Leggett (1992) 
these stars are in fact dM2 stars. We shall use in the future this calibration 
and therefore all our previous work on ``dM1" stars in this series of papers 
should be considered as papers referring to dM2 stars. 

Our principal goal in this series of papers is to constrain the Mass-Activity
Empirical Correlations by determining the radii of our sample stars as well as 
the level of magnetic activity in the Ca\,{\sc ii} and H$_{\alpha}$ spectral 
lines. Searching through databases at the European Southern Observatory (ESO) 
and Observatoire de Haute Provence (OHP), we identified spectra of about 800 
different stars which are suitable for our purposes. These spectra will be 
studied in Houdebine et al. (2019b, in preparation). The spectra of these 
datadases are completed by our own observations with the spectrgraphs SOPHIE 
(OHP) and NARVAL (Pic-Du-Midi). These observations include about 430 high 
resolution spectra of dwarfs never observed in the Ca\,{\sc ii} lines in 
high resolution spectroscopy before, from late-K to M5. 

Here, we focus on the determination of the effective temperature and radius 
for the stars in our samples. In the samples presented here, we do not include 
all the stars in the lists above except for the M6 and M7 samples for which 
all our stars are included in order to obtain better statistics. For our 
samples, we notably included all stars with determined metallicities. The 
final list of our samples of stars is provided in Table~2.

\section{Stellar parameters}

Here, we discuss the methods we use to derive $T_{eff}$ and radius for each of 
our samples of stars. Reliable estimates of the radii and masses are important 
if we wish (as in paper II) to determine reliable Mass-Activity Empirical 
Correlations. 

\subsection{Effective temperatures}

B12 used interferometric observations to determine with high precision (better 
than 5\% ) the diameters of 33 K and M dwarfs. They derived empirical 
correlations linking the effective temperature to broadband colours including 
(V-R) and (V-I) (from which we derive (R-I)). Their correlations were obtained 
for stars ranging in spectral type from K0 to M4. Similarly, M15 derived 
accurate stellar parameters, notably the radii of K7-M7 single stars with a 
precision of 2-5\%, as well as model-independent relations between $T_{eff}$ 
and broad-band colours. We show the empirical relationships between $T_{eff}$ 
and the (R-I)$_C$ colour in Fig.~1 for these two calibrations. We give the 
measurements of (R-I)$_C$ and $T_{eff}$ from B12 and M15 in Table~1. We can 
see in Fig.~1 that the two parameters are tightly correlated for dwarf stars 
ranging in spectral type from K0 to M7. We also show the calibration of 
$T_{eff}$ as a function of (R-I)$_C$ from Kenyon \& Hartmann (1995) in Fig.~1 
for comparison. We overplot a smoothing of the data of B12 and M15 (continuous
 line). In the present paper, we use the smoothed data (i.e. the continuous 
line in Fig. 1) to derive ``our" values of $T_{eff}$ for each of the objects 
in B12 and M15. Once we have derived ``our" value of $T_{eff}$ for each object 
in both of those papers, we then calculate the temperature differences between 
our values and those of B12 and M15. We find that the mean of the differences 
between ``our" values and those of B12 and M15 is only $\pm 3\sigma =$44.6~K 
for K dwarfs (23 stars) and $\pm 3\sigma =$39.4~K for M dwarfs (161 stars). 
These values demonstrate that the (R-I)$_C$ color is a reasonable  
temperature diagnostic for most K and M dwarfs.

In this regard, we note that the calibration of Kenyon and Hartmann (1995) 
agrees well with the measures of B12 and M15 up to (R-I)$_C$=1.3, but tends to 
overestimate $T_{eff}$ for later spectral types. For comparison between the 
different spectral sub-types used in this study, we overplot the (R-I)$_C$ 
domains of our samples of stars in Fig.~1 from mid-K to dM7.

Using the smoothed fit (continuous line in Fig. 1) to the calibrations of B12 
and M15, we have determined values of $T_{eff}$ from our compilations of 
(R-I)$_C$ measurements for all our target stars (see Tables~1 and 2). In 
Table~2, values of $T_{eff}$ which we have derived from our smoothed 
continuous line in Fig.~1 are listed in column 4.

We also aim in this study {\em to provide an extensive compilation of 
$T_{eff}$ compiled from the literature for our objects}. The results of this 
extensive search are listed in column 5 of Table~2 for all stars in our sample.

In order to have a complete compilation of previous measurements of $T_{eff}$, 
we queried all catalogs in the VizieR database (CDS, Strasbourg, France: 
\footnote{http://vizier.u-strasbg.fr/viz-bin/VizieR}). To this respect, we 
used the tutorial developed by Paletou \& Zolotukhin 
(2014\footnote{http://arxiv.org/abs/1408.7026}). The results of the queries provide 
us with thousands of measurements for several thousands of stars in all our 
target lists. However, the queries are carried out for all objects around 
the coordinates of the target stars. For instance, in 
case of binaries, most often, measurements of the two binary components are 
included in the compilations. Also, some other brighter or fainter stars (for 
instance white dwarfs) may be included. As a 
consequence, many spurious $T_{eff}$ measurements are included in our 
compilations. In order to be sure we have only the correct measurements of our 
target stars, we had to query separately many catalogs individually for many 
of our targets. Our compilation of temperatures includes all measurements 
regardless of the methods used to derive them.

The sources of the published effective temperatures for our K dwarf samples 
are: Blackwell \& 
Lynas-Gray (1998), Soubiran et al. (1998), Cenarro et al. (2001), Borde et al. 
(2002), Gray et al. (2003), Le Borgne et al. (2003), Wright et al. (2003), 
Yong \& Lambert (2003), Clem et al. (2004), Kovtyukh et al. (2004), Valenti \& 
Fischer (2005), Ammons et al. (2006), Casagrande et al. (2006), Gray et al. 
(2006), Masana et al. (2006), Sanchez-Blazquez et al. (2006), Sousa et al. 
(2006), Cenarro et al. (2007), Carney et al. (2008), Casagrande et al. (2008), 
Morales et al. (2008), Soubiran et al. (2008), Sousa et al. (2008), Cort\'es 
et al. (2009), da Silva et al. (2009), Jenkins et al. (2009), Mantega et al. 
(2009), Oenehag et al. (2009), Lafrasse et al. (2010), Soubiran et al. (2010), 
Valentini \& Munari (2010), Casagrande et al. (2011), Malyuto \& Shvelidze 
(2011), Prugniel et al. (2011), Wright et al. (2011), McDonald et al. (2012), 
Bermejo et al. (2013), Molenda-Zakowicz et al. (2013), Pace (2013), Stelzer et 
al. (2013), Tsantaki et al. (2013), Chen et al. (2014), Cottaar et al. (2014), 
Franchini et al. (2014), Gaidos et al. (2014), Munari et al. (2014), Kopytova 
et al. (2016).

For the M dwarfs, the effective temperatures come from the following authors: 
Silva \& Cornell (1992), Blackwell \& Lynas-Gray (1998), Cenarro et al. 
(2001), Malkan et al. (2002), Gray et al. (2003), Le Borgne et al. (2003), 
Wright et al. (2003), Allende Prieto et al. (2004), Clem et al. (2004), Valdes 
et al. (2004), Valenti \& Fischer (2005), Ammons et al. (2006), Butler et al. 
(2006), Gray et al. (2006),, Sanchez-Blazquez et al. (2006), Sousa et al. 
(2006), Cenarro et al. (2007), Schiavon (2007), Baines et al. (2008), 
Casagrande et al. (2008), Morales et al. (2008), Soubiran et al. (2008),  
Sousa et al. (2008), Jenkins et al. (2009), Schroeder et al. (2009), van Belle 
\& von Braun (2009), Brown (2010), Gazzano et al. (2010), Houdebine (2010), 
Lafrasse et al. (2010), Soubiran et al. (2010), Casagrande et al. (2011), 
Malyuto \& Shvelidze (2011), Prugniel et al. (2011), Wright et al. (2011), 
Christiansen et al. (2012), Houdebine (2012), Houdebine et al. (2012), Koleva 
\& Vazdekis (2012), McDonald et al. (2012), Rojas-Ayala et al. (2012), Bermejo 
et al. (2013), Cesetti et al. (2013), Lepine et al. (2013), Molenda-Zakowicz 
et al. (2013), Pace (2013), Rajpurohit et al. (2013), Stelzer et al. (2013), 
Gaidos et al. (2014), Loyd \& France (2014), Munari et al. (2014), Rajpurohit 
et al. (2014), Eker et al. (2015), Frasca et al. (2015), Mann et al. (2015), 
Newton et al. (2015).

We found from these authors that differences in $T_{eff}$ between different 
sources are commonly of the order of 100-200~K and may even exceed 400-500~K !
We found that these differences are the greatest among the mid-K, late-K, dM6 
and dM7 samples. They are somewhat smaller among the dM0-dM1, dM2, dM3, dM4 and
 dM5 samples. We found for instance that the $T_{eff}$ measures of Jenkins et 
al. (2009) and Lepine et al. (2013) are systematically underestimated. We found
 in general that the $T_{eff}$ are underestimated for M0-M5 dwarfs compared to 
our values derived from the (R-I)$_C$-$T_{eff}$ calibration of B12 and M15. 
Nevertheless, our values are in good agreement with those from Gaidos et al. 
(2014), and may even be slightly underestimated.

We computed the means of the temperature differences between our measures from 
(R-I)$_C$ and from the literature for all our samples. We find that in 
average, the $T_{eff}$ differences are 34.78~K, 41.40~K, 22.11~K, 29.81~K, 
25.50~K, 27.60~K, 32.08~K, 17.90~K and 54.36~K for our mid-K, late-K, dM0-dM1, 
dM2, dM3, dM4, dM5, dM6 and dM7 stellar samples respectively. We assigned 
these values as estimates of the $\pm 3\sigma$ uncertainty on our measures of 
$T_{eff}$ derived from (R-I)$_C$. We can see from these values 
that the $T_{eff}$ measurements from the literature are best determined for 
mid-K, dM3 and dM6 stars, but are worse determined for late-K, dM5 and dM7 
stars. We emphasize the relatively large uncertainty on the determinations of 
$T_{eff}$ for the dM7 stars (over $\pm 100$~K). We also would like to 
emphasize that the differences between our measures of $T_{eff}$ derived from 
(R-I)$_C$ and those of the literature directly depends on the number of 
literature measurements: i.e. the largest are the numbers of measures, the 
smallests are the differences with our measures. On the contrary when only one 
measure is available from the literature, the difference with our $T_{eff}$ 
derived from (R-I)$_C$ is generally large. The means of the temperature 
differences provide us with estimates of the $\pm 3\sigma$ uncertainty on our 
$T_{eff}$ determinations from (R-I)$_C$. We stress here that these 
uncertainties are relatively low (as low as $\pm 18$~K) compared to the 
typical $\pm 100$~K uncertainty claimed by most authors. This underlines that 
the (R-I)$_C$ colour is a reasonable effective temperature diagnostic for 
most K and M dwarfs, and that the calibrations we use in this study are 
reasonably reliable. We also stress that the (R-I)$_C$ colour give good 
effective temperature estimates even for sub-dwarfs (e.g. Gl 130, Gl 333, 
Gl 438, Gl 563.2A, Gl 563.2B, Gl 637 and Gl 817 in the dM2 sample). This 
highlights the relatively low sensitivity of the (R-I)$_C$ colour on the 
metallicity. We emphasize that the (R-I)$_C$ colour varies in time for many 
M dwarfs due to the presence of spots. This somewhat alters the precision of 
the (R-I)$_C$ measurements from the literature for active dwarfs. This is 
especially true for M6 and M7 objects that are most often very active (see 
Houdebine et al. 2019b, in preparation).

We give in Table~2 the mean temperatures (column 6) of the temperatures 
derived from (R-I)$_C$ (column 4) and the temperatures compiled from the 
literature (column 5). We also give in Table~2 the uncertainties on our final
 $T_{eff}$ measures as the differences between $T_{eff}$ derived from (R-I)$_C$
 and $T_{eff}$ from the literature. When this value is not available (no 
measures from the literature), we assigned the uncertainty as the mean of the 
temperature differences.
 
\begin{figure*} 
\vspace{-0.2cm}
\begin{centering}
\hspace{-1.5cm}
\includegraphics[width=14cm,angle=+90]{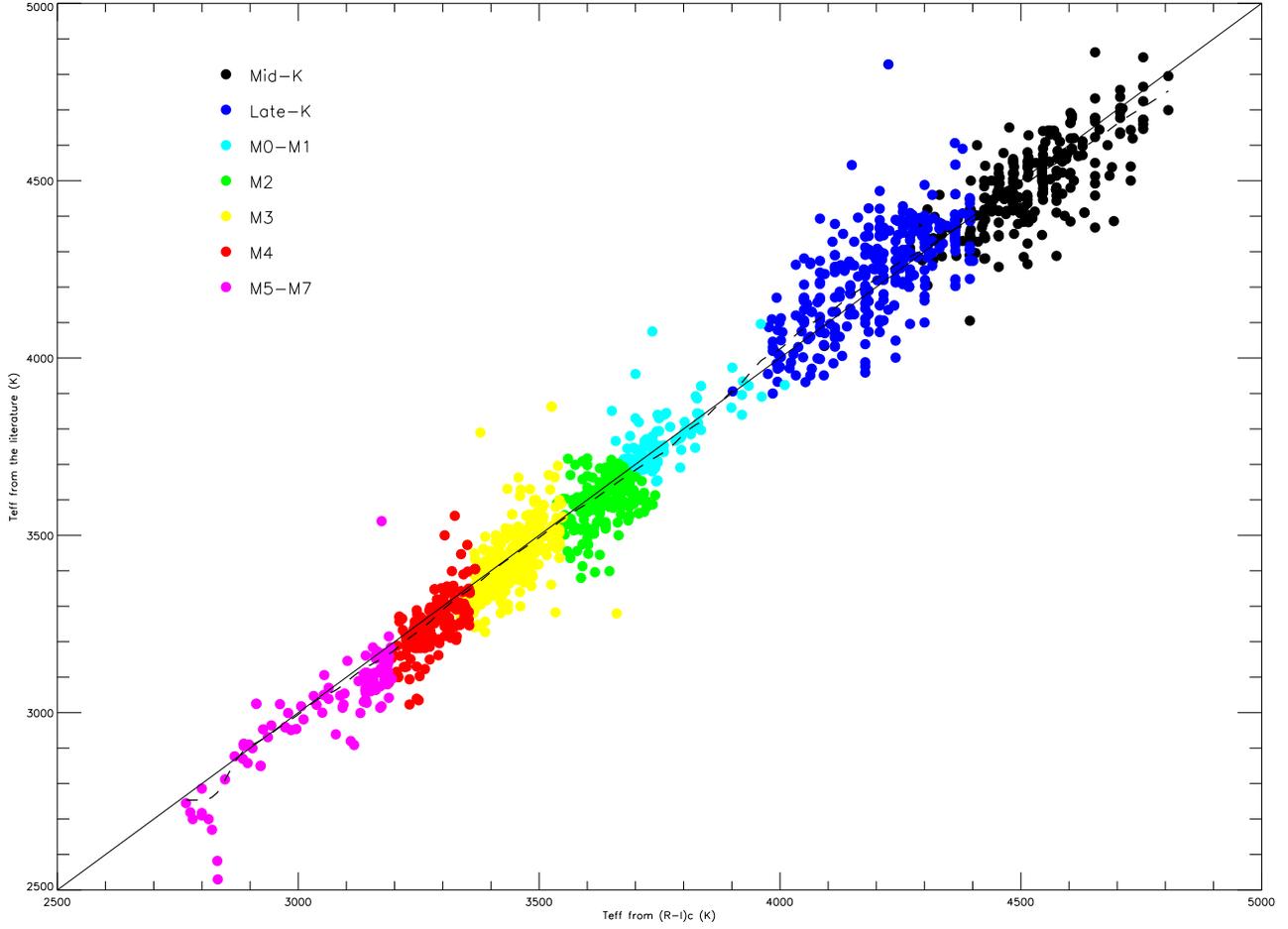}
\vspace{-0.5cm}
\end{centering}
\caption[]{ Comparison between two sets of values of $T_{eff}$: (i) 
values taken from the literature; and (ii) values which have been derived 
from the (R-I)$_C$/$T_{eff}$ correlations which have been derived by  M15 
and B12. Continuous line: the correlation which would exist if there was 
an exact one-to-one correspondence between the two sets of $T_{eff}$. 
Dashed line: the best fit correlation when the data are smoothed by a 
Gaussian of FWHM 30~K. Note that there exists a good overall agreement 
between the smoothed data and the expectations of a one-to-one 
correspondence.
}
\end{figure*}

\subsection{Effective temperatures: Systematic errors}

In order to better estimate the agreement between our final $T_{eff}$ and the 
$T_{eff}$ from previous authors, we plot these two parameters in Fig.~2. 
In Fig.~2, the continuous (solid) line indicates the correlation which 
would exist if there was perfect correspondence between the two approaches 
we have used to evaluate $T_{eff}$: (i) from the literature, and (ii) from 
the color-temperature relationships obtained by B12 and M15. It can be seen 
that the agreement between the datasets is rather good. 
The scatter of the measures around the mean curve is 
typically lower than 100~K at all spectral subtypes. We note that this scatter 
is lower for instance for mid-K dwarfs than for late-K dwarfs. This better 
agreement among mid-K dwarfs is due mostly to the larger number of $T_{eff}$ 
measurements from the literature for these stars. On the other hand, in 
general we have fewer estimates of $T_{eff}$ from the literature for late-K 
dwarfs. We again stress that a large part of the scatter observed in Fig.~2 is 
due to few measures of $T_{eff}$ from the literature. When large numbers of 
measures of $T_{eff}$ are available from the literature (e.g. for the Gliese 
stars), the agreement with our $T_{eff}$ derived from the (R-I)$_C$ color is 
generally good. We also note that the scatter is lower among M dwarfs compared 
to late-K dwarfs and especially for M4 to M7 stars.

The data in Figure 2 show that, for late-K dwarfs, when we compare the 
$T_{eff}$ values obtained by the two methods, the literature values of 
$T_{eff}$ are on average {\it larger} by about 6\% than the values of 
$T_{eff}$ which are obtained from the B12 calibration. In contrast to such 
behavior, the results in Figure 2 indicate that for mid-K dwarfs, we find 
that the literature values of $T_{eff}$ are on average {\it smaller} by 
about 9\% than the B12 values. Also for M dwarfs, the behavior is similar 
to that for the mid-K stars, i.e. the literature values of $T_{eff}$ are 
{\it smaller} by typically 5\%  than the B12 values. We note that Mann et 
al. (2013) also reported that their effective temperatures were slightly 
larger than those from other authors in the literature by about 50~K. M15 
found that their estimated $T_{eff}$ were typically 100~K warmer than 
those from the work of Casagrande et al. (2008). On the other hand, Mann 
et al (2013) reported on a good agreement between their $T_{eff}$ estimates  
and those of Rojas-Ayala et al. (2012). During the course of our data 
compilation, it is worth noting that we found good agreement between our 
$T_{eff}$ values derived from (R-I)$_C$ and the recent estimates of 
$T_{eff}$ reported by Gaidos et al. (2014).

As an example of a compilation of temperatures from the literature, we 
list in Table~3 the temperatures obtained by 6 different teams of authors 
for various sub-sets of the stars which overlap with those in our dM4 
sample. For the 6 teams as listed from left to right in Table 3, the 
number O(i) of stars which overlap with our sample are O(i) = 38, 30, 
11, 32, 60, and 111 respectively: the largest overlap (111) is with the 
team of Gaidos et al (2014), while the smallest overlap (11) is with the 
team of Wright et al (2011). We list, at the bottom of Table~3, the mean 
temperatures obtained by those 6 teams for their respective sub-set of dM4 
stars.  We also list, for each team (i=1~6) the mean temperatures which we 
have obtained for the dM4 stars in our sample (numbering O(i)) which 
overlap with the stars observed by Team (i). We also list the mean 
temperature differences between our values and those of the 6 authors: 
the differences are listed in absolute terms (in degrees K) and also as 
a percentage. These figures highlight the systematic differences between 
the results obtained by means of our approach and those obtained by 6 
other sets of authors. 

We note that, in the case of two of the 6 teams, there is very good agreement, 
$\pm$0.3\% on average, between our estimates of $T_{eff}$ and theirs: 
these are the teams of Gaidos et al. (2014) and of Stelzer et al. 
(2013). To understand why there is such a good overlap between ourselves 
and Gaidos et al. (2014), we note that Gaidos et al. followed the procedure 
of Mann et al. (2013) in order to estimate $T_{eff}$, $R_{*}$ and luminosity 
$L_{*}$ for their sample. Gaidos et al. first determined $T_{eff}$ by 
finding the best-fitting model stellar spectrum, and then they used the 
best-fitting temperature in empirical relations to determine the other 
stellar parameters. This procedure was calibrated against nearby stars with 
known radii, distances and bolometric fluxes (Boyajian et al. 2012). 
Therefore, their method is essentially the same as that used by M15: since 
the approach we adopt in the present paper also relies in part on M15, it 
is perhaps not too surprising that we have found a very good agreement with 
the results obtained by Gaidos et al (2014). Moreover, the overlap of stars 
between our sample and the sample of Gaidos et al is the largest (111 stars), 
and this may also contribute to improving the agreement between our results.

On the other hand, the good agreement that we have found with the results 
of Stelzer et al. (2013) cannot be attributed to similarity of approach. 
In fact, Stelzer et al. (2013) used a completely different technique based 
on the spectral type calibration reported by L\'epine et al. (2012). The 
latter calibration is based on the (V-J) color, and is therefore quite 
distinct from the method (based on the $(R-I)_C$ color) which we have 
used in the present paper. Once Stelzer et al. had determined the 
spectral type for each star, they then used the temperature scale 
reported by Bessell (1991) and Mohanti \& Basri (2003) in order to derive 
$T_{eff}$. In view of the difficulties associated with assigning a 
precise spectral type and then also converting to a temperature, this 
method is expected to be subject to several uncertainties. Nevertheless, 
as shown in Table 3, it is encouraging to see that we have actually found 
good overall agreement between the Stelzer et al. temperatures and ours. 
And we note that the overlap between our sample of dM4 stars and those 
of Stelzer et al. is only 32: thus, we cannot claim that the largeness 
of the example might be helping to bring our samples into agreement.
 
Turning now to the results of a third team, we note that Morales et al. 
(2008) have used the TiO5 index to derive spectral types for their sample 
of M dwarfs: the TiO5 index is based on the strongest TiO feature in M 
dwarf spectra, with a bandhead at 7050\AA . Effective temperatures were 
computed using the spectral-type temperature correlation reported by 
Bessel (1991). Morales et al. used an iterative procedure to ensure that 
their $T_{eff}$ values were consistent with the bolometric correction 
in the K band $BC_{K}$. We find (see Table 3) that the values of 
$T_{eff}$ obtained by Morales et al. for M4 dwarfs lie about 84~K below 
our own measures for the dM4 sample. We believe that this significant 
difference in $T_{eff}$ values can be attributed to the presence of a 
systematic difference between the calibration of Bessel (1991, see also 
Leggett et al. 1996), and the calibration of M15.

The team of Jenkins et al. (2009) used the (V-K$_S$)-$T_{eff}$ 
relation reported by Casagrande et al. (2008). The latter 
investigators claimed a typical internal uncertainty of $\pm$17~K. 
In preparation for using the Casagrande et al relation, Jenkins et al. 
first computed the absolute V and K magnitudes. The V magnitudes 
were taken from SIMBAD, while the $K_S$ magnitudes were taken from 
the 2MASS catalog (Skrutskie et al. 2006). We find that, on average, 
the $T_{eff}$ values obtained by Jenkins et al. are lower than ours 
by about 212~K. This figure is larger than the differences we have 
found relative to those of other teams. For this reason, we have 
rejected some of the Jenkins et al results in cases where the 
difference was significantly larger than the mean $T_{eff}$ values 
obtained by ourselves and by other teams. 

The team of Wright et al. (2011) have used a combination of techniques 
in order to estimate the temperatures for their targets. For most 
stars, they adopted two methods, both involving isochrones from 
Siess et al. (2000): (i) for cluster stars, they combined the 
isochrones with an estimate of the age of the appropriate cluster; 
(ii) for field stars, they assumed an age of 1 Gyr. For the field 
stars with unknown distances, the (V-K) color led them to derive 
a temperature using the relationships of Casagrande et al. (2008) 
for M dwarfs, and those of Casagrande et al. (2010) for FGK stars. 
Their results for M4 dwarfs are found to lie about 171~K hotter than 
our measures (+5.19\% ). The study by the Wright et al. (2011) team 
is the only one of the six teams which clearly overestimates the 
temperatures in mid-M dwarfs compared to the tabulation of Mann et 
al. (2015). We note that the sample of Wright et al (2011) has the 
smallest overlap with our sample of dM4 stars: only 11 stars 
contribute to the overlap. This smallness of overlap may serve to 
enhance the difference in the mean values of $T_{eff}$.  

Finally, the team of L\'epine et al. (2013) have obtained results for 
the most complete survey of M dwarfs in the northern sky. They 
determined the stellar parameters using a complex method based on 
fitting stellar atmosphere models from Allard et al. (2011). They fitted 
their spectra in the wavelength range 5600-9000\AA , but excluding the 
problematic TiO bands between 6400 and 6600\AA . Between the average 
values of $T_{eff}$ obtained by of L\~epine et al. and the averages 
which we have obtained, there is a systematic difference of -91~K (-2.78\% ).  

By taking a grand average of the 6 mean differences in $T_{eff}$ in 
Table 3, we find a value of -36 deg. K. Compared to the typical $T_{eff}$ 
values listed in Table 3, this grand average amounts to a fractional 
error of about 1\% .  

As we saw above, the rather good agreement between our values of $T_{eff}$ 
and those in the literature may be considered as somewhat surprising in 
view of the systematic differences between the various approaches adopted 
by the different authors. We believe that the sources of these systematic 
differences may include (but are not limited to) the following: missing 
opacities, the LTE assumption in the models, and metallicity effects on 
the (R-I)$_C$ color. Differences of typically 100~K or 200~K between 
different authors are common in our compilation of effective temperatures, 
mainly because of systematic errors. However, it appears that on average, 
these differences cancel one another to a greater or lesser extent, 
thereby yielding values which are not so far removed from our calibration 
based on the work of B12 and M15. Without doubt, we can assert that the 
best reference is the work of B12. 

It is true that we have found some discrepancies between the literature 
values of $T_{eff}$ and our values in the mean mid-K and late-K 
subsamples. These discrepancies may point to systematic errors in the 
methods used to derive values of $T_{eff}$ in the literature. Because 
our compilation includes works by authors who have relied on many 
different techniques, we believe it is beyond the scope of this paper 
to discuss in detail all possible sources of uncertainties in detail. 
We only can say that in average, systematic errors for individual stars 
are typically less than 10\% , while the errors which appear in the 
grand average value of $T_{eff}$ are of order 1\% .  In the case of 
the K dwarfs, we find that there are systematic errors of about 5\% 
for the $T_{eff}$ values which we derive for K dwarfs: we find that 
the sign of the systematic error is positive for late-K stars and 
negative for mid-K stars.

For M dwarfs, our results for $T_{eff}$ are based mostly on the 
approach which was developed by M15. The uncertainties associated with 
their results have already been discussed in the Introduction. The 
M15 work seems to be one of the most precise studies so far, 
essentially because it includes results from interferometry. Since 
we rely heavily on M15, we recognize that whatever sources of error 
contribute to the results in M15 also apply to our results. Moreover, 
we find that our $T_{eff}$ values are systematically larger than the 
literature values for most authors, with the notable exception of 
Gaidos et al. (2014). We have found that the systematic differences 
in $T_{eff}$ are of order 5\% for M4 and M5 dwarfs. However, the 
percentage errors are larger for the less well known M7 dwarfs. On 
the other hand, we have found good agreement for the M2 and M3 
subsamples. 

In summary, our results for $T_{eff}$ include systematic errors of 
about 5\% overestimates, in addition to which we should allow for 
the uncertainties in the results of M15 (see Introduction).

In Table~2, the errors given in column 6 are the differences 
between the estimated effective temperature and the solid line 
curve in Fig.~1. This definition means that the difference is 
between our $T_{eff}$ value and the mean of the values of 
$T_{eff}$ in the literature. However, it is important to note 
that errors on the literature values of $T_{eff}$ may be as 
large as 200~K ! Therefore the mean errors on our samples given 
above are only statistical measures of the mean of the scatter 
relative to the straight line in Fig.~2. It is difficult at 
this stage to estimate absolute errors for our $T_{eff}$ values 
since many different sources contribute to the errors.

Another additional source of systematic error is the effect of 
[M/H] on the (R-I)$_C$ color. Leggett (1992) found that for a 
given (R-I)$_C$ color, there exists one spectral subtype 
difference between Young Disk and Halo M dwarfs. In other 
words, for a given (R-I)$_C$, low metallicity M dwarfs tend to 
be cooler than solar metallicity dwarfs. The consequence is 
that our values of $T_{eff}$ derived from (R-I)$_C$ tend to be 
cooler for subdwarfs. This is one reason why we have compiled 
metallicities for our targets. Even if this systematic trend 
has been well established by other authors, we find (see 
Table~2) that in general, for the vast majority of our 
subdwarfs the difference between our $T_{eff}$ values and the 
values in the literature remains small, except for very low 
metallicity subdwarfs (e.g. the sdM2 VB 12).

Moreover, it is also a fact that the presence of surface 
inhomogeneities on the surface of an M dwarf (especially on a dMe 
star) can have an effect on the value of $T_{eff}$ which is 
measured for such a star. This effect applies to our data, as 
well as values of $T_{eff}$ which appear in the literature. 
Systematic errors due to this effect can hardly be determined 
with any reliability unless full investigations of the spectral 
variations in time have been established. We note also that 
there are some variations of the (R-I)$_C$ colors in our 
compilations, which may yield uncertainties up to about 100~K 
for certain objects. For the best known objects, we have sometimes 
found several reported measurements of (R-I)$_C$, and for such 
objects, we use the mean. Such variability in the actual color 
of a particular star can be responsible for a significant part 
of the scatter in Fig.~2,

\begin{figure*} 
\vspace{-0.2cm}
\begin{centering}
\hspace{-1.5cm}
\includegraphics[width=14cm,angle=+90]{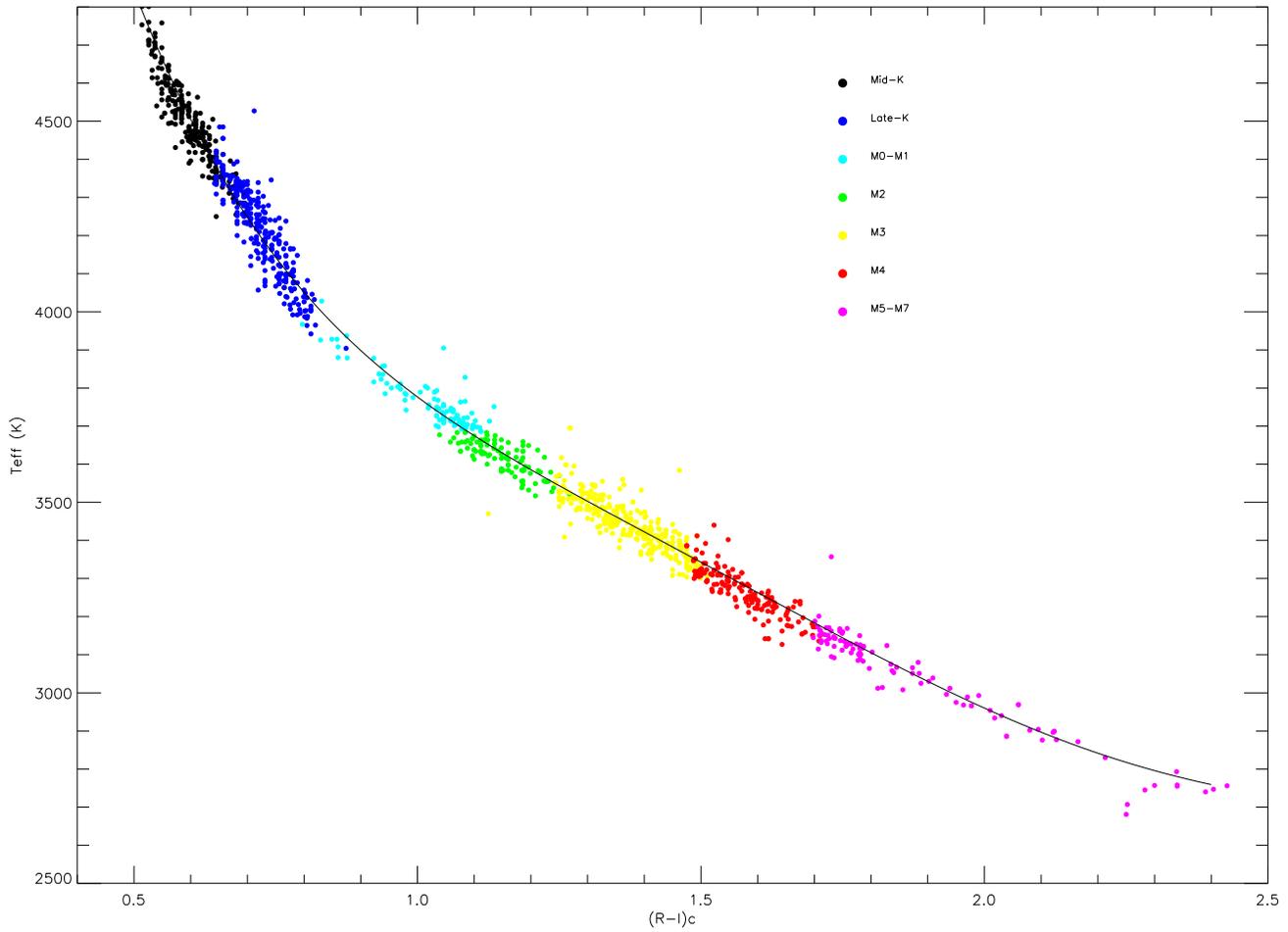}
\vspace{-0.5cm}
\end{centering}
\caption[]{Values of $T_{eff}$ compiled in the present study as a 
function of (R-I)$_C$ for our complete samples of stars. Solid line: The 
heteroscedastic least-square fit to the data (see text).}
\end{figure*}

We show in Fig.~3 the values of our final effective temperatures  as a 
function of (R-I)$_C$ for our complete samples of stars. One can see in this 
figure that the correlation between $T_{eff}$ and (R-I)$_C$ is good. This 
confirms that (R-I)$_C$ is a good first guess effective temperature diagnostic 
for K and M dwarfs. We show the heteroscedastic least-square fit to the data 
as the solid line. Weights are proportional to the inverse of the error on 
the final temperature determination. This least square fit of order 5 yields:

\begin{eqnarray}
T_{eff}/3500 = -2.3365\times (R-I)_{C}^{5} \nonumber \\
 + 4.1056\times (R-I)_{C}^{4} \nonumber \\
-3.7582\times (R-I)_{C}^{3}  \nonumber \\
+ 2.4905\times (R-I)_{C}^{2}\nonumber \\
-0.06876\times (R-I)_{C} + 0.646327
\end{eqnarray}

The $\chi^2$ is $3.56\times 10^{-6}$ for this fit, which is good. Most stars 
in our sample lie within $\pm$100~K of the least squares fit. This fit could 
be used as a complementary calibration of $T_{eff}$ as a function of (R-I)$_C$ 
for K and M dwarfs as it is based on both the work of B12 and M15, and our 
complete compilation of $T_{eff}$ from the literature.

\subsection{Stellar radii}

In order to derive radii for our samples of stars, we used the classical 
formula e.g. Lang 1980);

\begin{equation}
M_{v}+BC_{V} = 42.36 - 5\times log(\frac{R_{*}}{R_{\odot}}) - 10\times 
log(T_{eff}),
\end{equation}

\noindent
where symbols take their usual meaning. We used the $BC_{V}$ calibration as a 
function of $T_{eff}$ from Lejeune et al. (1998). We assumed that their 
tabulation yields an uncertainty of about 10\% in $BC_{V}$. Using Eq. (2), we 
obtained the radii for our samples of stars as listed in Table 2. The errors 
on the radii in Table~2 were computed taking into account the combination of 
the error on the absolute magnitude, an assumed error of 0.02 magnitudes on 
the (R-I)$_C$ colour, an error of 0.005 magnitudes when $V$ is known to 3 
decimals, 0.02 magnitudes when $V$ is known to 2 decimals, and 0.2 magnitudes 
when $V$ is known to 1 decimal. We also took into account the errors on 
$T_{eff}$. 

Our source for the parallaxes comes mostly from the recent GAIA DR2 
compilation (GAIA collaboration 2016, 2018). Other sources are Jenkins (1952), 
Gliese \& Jahreiss (1991), Van Altena et al. (1995), Salim \& Gould (2003), 
Costa et al. (2005), van Leeuwen (2007), Subasavage et al. (2009), Dieterich 
et al. (2014), Lurie et al. (2014). The errors on the parallaxes were included 
in our calculations of the $V$-band absolute magnitude $M_V$. We found that 
the GAIA DR2 data allows a tremendous improvement on the precision of the 
stellar radii especially over the older data of Gliese \& Jahreiss (1991). 
In some instances we found radii different by a factor of 2 or 3 between 
these two sources. We also found some significant differences with the 
Hipparcos parrallaxes in many instances, notably for faint Hipparcos sources.

\begin{figure*} 
\vspace{-0.1cm}
\begin{centering}
\hspace{-1.5cm}
\includegraphics[width=14cm,angle=+90]{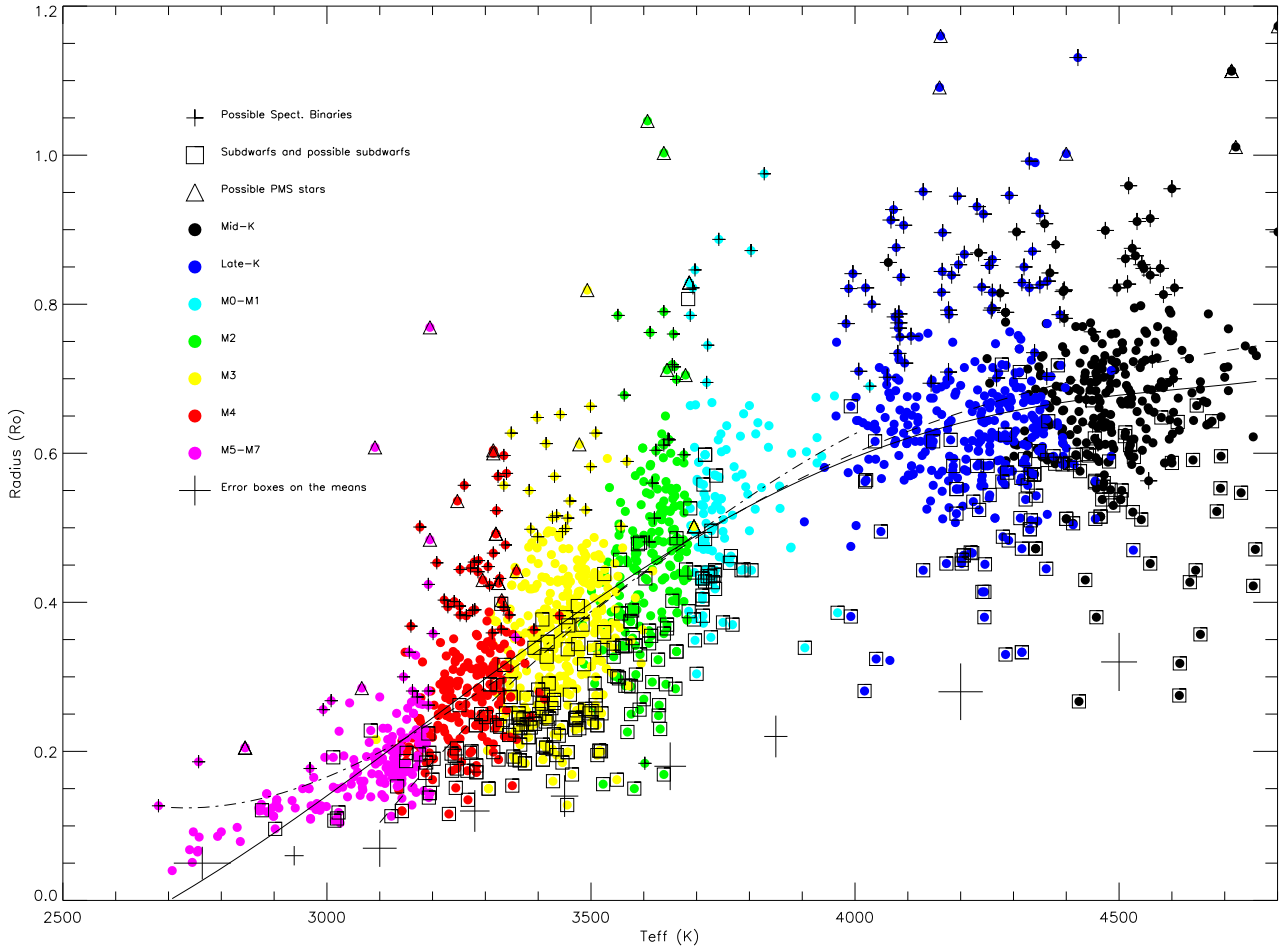}
\vspace{-0.5cm}
\end{centering}
\caption[]{Stellar radius as a function of $T_{eff}$ for our samples of K and 
M dwarfs. We show the heteroscedastic least square fit as the continuous 
line. The dashed line shows the fit from Boyajian et al. (2012). The dot-dashed
 line represents the fit determined by Mann et al. (2015) for their sample of 
stars. The large scatter which is present in this figure is due (in large part)
to the effects of different values of metallicity among the individual stars 
in any sample.}
\end{figure*}

We show the stellar radii as a function of $T_{eff}$ in Fig.~4 for all of the 
(1910) stars in our samples. We show the weighted least squares fit as 
the continuous line for stars restricted to $-0.5\leq [M/H]\leq +0.5$. The 
dashed line represents the polynomial fit determined 
by B12 for their sample of (33) K and M stars. We also show as the dot-dashed 
line, the polynomial fit obtained by M15, for their sample of (183) K and M 
dwarfs. For our present samples of stars we obtained the following weighted 
polynomial fit:

\begin{eqnarray}
R_{*} = 2.8102\times X^{6}-7.5796\times X^{5}+1.2666\times X^{4} \nonumber \\
+9.0658\times X^{3}-2.8046\times X^{2} -3.9200\times X \nonumber \\
 + 1.5606
\end{eqnarray}

where $X= T_{eff}/3500.$. 
The weights 
on each data point were assigned as the inverse of the errors on the radii.

We overplot for each of our spectral sub-type domain the mean of the errors on 
$T_{eff}$ as well as the mean of the errors on $R_{*}$. We show these error 
bars in the lower part of Fig.~4.

The first striking characteristic of the data represented in Fig.~4 is the 
very large scatter among the radii for a given effective temperature. This 
scatter can be attributed partly to uncertainties on $T_{eff}$ and $R_{*}$ as 
can be seen on the means of the error bars plotted in Fig.~4. However, the 
scatter is still significantly larger than these typical error bars, notably 
for the earliest K dwarfs in our sample as well as the early M dwarfs. 
Such a large scatter in the radii was also observed among M dwarfs by Mann et 
al. (2015). They showed that the differences between radii at a given effective
temperature is mainly due to differences in metallicities: stars with large 
radii are metal rich and stars with small radii are metal poor (e.g. 
subdwarfs). Houdebine (2008) found that there exists an empirical correlation 
between stellar radius and metallicity among his sample of M2 dwarfs. Later, 
Houdebine et al. (2016) found also analogous empirical correlations among 
their samples of late-K, M3 and M4 dwarfs. Although it appeared in this later 
study that for M4 dwarfs, the dependency of the stellar radius on the 
metallicity is somewhat smaller than in say, late-K or M2 dwarfs, it seems 
that both parameters do correlate among K and M dwarfs. The smallest 
dependency of late M dwarf radii on metallicity may explain why we observe a 
smaller scatter in Fig.~4 for spectral types equal or greater than M5, 
compared to earlier spectral type M dwarfs. Note also, that there are probably 
some (so far) unidentified binary stars in our samples (stars with abnormally 
large radii), and that our samples also contain subdwarfs.

We find a generally good agreement between our fit and the fits of B12 and M15 
from late-K to M3. However, we note some differences for early-K stars and 
stars later than M3. For M4 and particularly M5, the B12 curve lies by 40\% or 
more below our fit and the fit of M15. The fit from M15 agrees well with our 
fit from M1 to M5 stars, which is not surprising since for M dwarfs our 
temperature determinations depends for 50\% on their calibration, and the 
other 50\% depend on our literature compilation. For lower effective 
temperatures, the fit of M15 lies above our fit by up to about 70\%. 
Undoubtedly our sample is larger than that of M15 by a factor of 12 and larger 
than the B12 sample by a factor of 80. If we trust that the B12 and M15 
temperature determinations give "accurate" values, and that our final 
temperature determinations rely on a compilation from all other authors by up 
to 50\% , we may be tempted, considering our much larger sample, that our 
fit should yield a better description of the temprature-radius plane. But this 
is somewhat open to discussion since the errors on the temperatures may still 
be large.

We note that our fit do not reproduce well the radii of M6 and M7 dwarfs. This 
is due to the fact that errors on the radii for those stars are proportionally 
larger than those in M4 and M5 dwarfs, and as a result the weights assigned to 
M6 and M7 dwarfs are lower. The much larger samples of M4 and M5 dwarfs 
"force" the slope of the fit at M6 and M7 to be larger than expected.

The GAIA DR2 parallaxes allow a major improvement on the precision of the 
stellar radii of our nearby stars. Before, uncertainties on the parallaxes 
accounted for a large part of the uncertainties on the radii, especially for 
the faint M dwarfs. In the present study, we find that the largest 
uncertainty now arise from the uncertainty of the effective temperatures. 
The better precision on the radii now allows us to isolate possible 
spectroscopic binaries or stars with large metallicities which stand 
significantly above the main trend, and probable low metallicity stars 
(probable subdwarfs) that stand significantly below the main trend in the 
radius-temperature plane. The comparison of the radii of those stars with 
those of stars in our samples with $[M/H]<-0.5$ or $[M/H]>+0.5$ (from our 
compilation of [M/H] from the literature), indicates the stars with possible 
metallicities below -0.5 or above +0.5. We show the known and possible metal 
poor stars as squares in Fig.~4. We also show in this figure, the possible 
metal rich and/or spectroscopic binaries which display abnormally large radii. 
We indentified 211 possible low metallicity dwarfs, 184 metal rich and/or 
spectroscopic binaries. We also found 30 possible PMS dwarfs which have 
abnormally large radii and are known to be neither metal rich nor 
spectroscopic binaries according to our high resolution spectra. From the 
analysis of our high resolution spectra, we have identified 21 new 
spectroscopic binaries and 1 new triple system.

Our correlation between radius and $T_{eff}$ can be compared to the 
Hertzsprung-Russell diagrams for nearby stars from the GAIA Collaboration 
(2018b). They note a steepening of the slope at about $M_G \sim 11$ in 
their Fig.~6. Our data do not show a clear evidence for a similar increase 
in the slope in Fig.~4. The Radius-$T_{eff}$ diagrams in Rabus et al. 
(2019) show that there are two possible regimes in their diagrams and they 
propose that their results are possible evidence for a change between 
partially convective stars and fully convective stars (The Transition To 
Complete Convection, TTCC). They locate this transition at about 
$T_{eff}\sim 3300$K. This result agrees with the finding of Houdebine et 
al. (2017) who found that the RACs undergo important changes between dM2 
and dM3 where they located the TTCC. This finding suggests that important 
changes occur in the dynamo mechanisms at the TTCC.

\section{Conclusion}

From our compilations of (R-I)$_C$ measures from the literature, we report on 
new derivations of effective temperatures for 1910 K and M dwarfs, based on 
the calibrations of B12 and M15. We have also compiled previously published 
effective temperatures for most stars in our sample. We derived our final 
effective temperature estimates by computing the mean of the effective 
temperatures derived from the (R-I)$_C$ colour and those from the literature. 
We find in general a good agreement between our temperatures derived from 
(R-I)$_C$ and the published effective temperatures (see Table~2).

We derive a new heteroscedastic polynomial fit to the $T_{eff}$ versus 
(R-I)$_C$ correlation. Our correlation is based both on the B12 and M15 
calibrations and our compilation of $T_{eff}$ from the litterature for 
thousands of measurements. We find a rather good (R-I)$_C$-$T_{eff}$ 
correlation in agreement with the results from previous authors. 

The recent release of the GAIA DR2 data allows us to compute the stellar radii 
for most of the targets in our stellar sample. We now find that the largest 
uncertainties on the radii arise from the uncertainties on the effective 
temperatures. We are currently carrying out a long term observing program 
dedicated to the measurements of chromospheric lines in M and K dwarfs. So far,
we have identified 21 new spectroscopic binaries and 1 triple system. We find 
that the precision on the stellar radii may allow to identify new low 
metallicity stars as well as new high metallicity and/or spectroscopic 
binaries. Our data suggest that 425 such new objects may be present in our 
data sets.

\section*{Acknowledgements}
This research has made use of the SIMBAD database, operated at CDS, Strasbourg,
France. DJM is supported in part by the NASA Space Grant program. This 
research made use of Astropy\footnote{http://www.astropy.org/ and 
http://astroquery.readthedocs.org/en/latest/}, a community-developed core 
Python package for Astronomy (Astropy Collaboration, 2013). This research has 
made use of the VizieR catalogue access tool, CDS, Strasbourg, France. The 
original description of the VizieR service was published in A\&AS 143, 23.

This work has made use of data from the European Space Agency (ESA) mission
{\it Gaia} \\
(\url{https://www.cosmos.esa.int/gaia}), processed by the {\it Gaia}
Data Processing and Analysis Consortium (DPAC,\\
\url{https://www.cosmos.esa.int/web/gaia/dpac/consortium}). Funding for the 
DPAC has been provided by national institutions, in particular the institutions
participating in the {\it Gaia} Multilateral Agreement.

\onecolumn

\scriptsize
\begin{longtable}{|lccccc|}
\caption[ ]{Effective temperatures and (R-I)$_{C}$ colors for K and M dwarfs. 
In Column 3, we list the $T_{eff}$ values we have obtained in the present 
paper using our (R-I)$_{C}$-$T_{eff}$ calibration. In Columns 4 and 5, we list 
the $T_{eff}$ values reported by M15 and by B12 respectively. } \\

\hline       &           &               &           &               &        \\
Star         &(R-I)$_{C}$& $T_{eff}$\footnotemark[1] & $T_{eff}$\footnotemark[2] &$T_{eff}$\footnotemark[3]& $T_{eff}$ Diff  \\
             &   (mag)   &     (K)       &     (K)   &     (K)       &  (K)   \\
             &           &               &           &               &        \\
\endfirsthead
\caption{continued.}\\

\hline       &           &               &           &               &        \\
Star         &(R-I)$_{c}$& $T_{eff}$\footnotemark[1] & $T_{eff}$\footnotemark[2] &$T_{eff}$\footnotemark[3]& $T_{eff}$ Diff  \\
             &   (mag)   &     (K)       &     (K)   &     (K)       &  (K)   \\
             &           &               &           &               &        \\
\endhead  
\hline
\endfoot

             &           &               &           &               &        \\
Gl 702A      &   -       &       -       &     -     &       5407    &    -   \\
Gl 75        &  0.39     &      5346     &     -     &       5398    &   +52  \\
Gl 631       &  0.39     &      5346     &     -     &       5337    &    -9  \\
Gl 53A       &  0.41     &      5259     &     -     &       5348    &   +89  \\
Gl 764       &  0.41     &      5259     &     -     &       5246    &   -13  \\
Gl 166A      &  0.45     &      5085     &     -     &       5202    &  +117  \\
Gl 144       &  0.46     &      5041     &     -     &       5077    &   +36  \\
\hline 
\end{longtable}
\normalsize
\footnotetext[1]{Effective temperature from the (R-I)$_C$ color/$T_{eff}$ calibration}
\footnotetext[2]{Effective temperature from Mann et al. 2015}
\footnotetext[3]{Effective temperature from Boyajian et al. 2012}

\scriptsize
\begin{landscape}
\begin{longtable}{|lcccccccccccc|}
\caption[ ]{Parameters of our sample of mid-K, late-K stars, M0-M1, 
M2, M3, M4, M5, M6 and M7 dwarfs. We also indicate the number of 
high resolution spectra for the stars that have observed.} \\

\hline    &       &       &               &           &               &        &                &                 &                &         &         &         \\
Star      &   V   &(R-I)$_{C}$& $T_{eff}$\footnotemark[1] & $T_{eff}$\footnotemark[2] &$T_{eff}\pm 3\sigma$\footnotemark[3]&Spect.& $\pi$ & $M_{V}$ &  $R_{\star}$  & [M/H] &Hi. Res. & Comments \\
          & (mag) & (mag) &     (K)       &     (K)   &     (K)       & Type   &     (m'')      &    (mag)        &  ($R_{\odot}$) &         & Spectra &         \\
          &       &       &               &           &               &        &                &                 &                &         &         &         \\
\endfirsthead
\caption{continued.}\\

\hline    &       &       &               &           &               &        &                &                 &                &         &         &         \\
Star      &   V   &(R-I)$_{C}$& $T_{eff}$\footnotemark[1] & $T_{eff}$\footnotemark[2] &$T_{eff}\pm 3\sigma$\footnotemark[3]&Spect.& $\pi$ & $M_{V}$ & $R_{\star}$  & [M/H] &Hi. Res. &  Comments \\
          & (mag) & (mag) &     (K)       &     (K)   &     (K)       & Type   &     (m'')      &    (mag)        &  ($R_{\odot}$) &         & Spectra &         \\
          &       &       &               &           &               &        &                &                 &                &         &         &         \\
\endhead  
\hline
\endfoot

          &       &       &        &           &               &        &                   &                 &                 &         &         &         \\
          &       &       &      \multicolumn{4}{c}{Mid-K sample}                           &                 &                 &         &         &         \\
          &       &       &        &           &               &        &                   &                 &                 &         &         &         \\
GJ 1139   & 9.567 & 0.656 &  4366  &    4391   &  4378$\pm$13  & dK5.4  & 43.4531$\pm$0.0293& 7.757$\pm$0.021 & 0.592$\pm$0.016 &  -0.33  &         &         \\
GJ 1176   & 9.225 & 0.620 &  4456  &    4509   &  4482$\pm$27  & dK4.3  & 40.6674$\pm$0.0433& 7.271$\pm$0.022 & 0.686$\pm$0.025 &  +0.07  &    35   &         \\
GJ 1189   & 9.30  & 0.566 &  4591  &    4577   &  4584$\pm$7   & dK3.6  & 30.6458$\pm$0.0552& 6.732$\pm$0.024 & 0.813$\pm$0.018 &  +0.04  &         & Binary? \\
GJ 1190   & 9.839 & 0.644 &  4396  &    4333   &  4364$\pm$31  &sdK5.4  & 38.8659$\pm$0.0589& 7.787$\pm$0.023 & 0.591$\pm$0.024 &  -0.95  &     5   &         \\
*GJ 1257  & 9.811 & 0.694 &  4270  &    4310   &  4340$\pm$30  & dK5.6  & 41.3937$\pm$0.0643& 7.896$\pm$0.023 & 0.572$\pm$0.023 &  -0.23  &         & Subdwarf? \\
\hline 
\end{longtable}
\end{landscape}
\normalsize
\footnotetext[1]{From the (R-I)$_C$ color.}
\footnotetext[2]{Compiled from the literature.}
\footnotetext[3]{Mean effective temperature.}

\scriptsize
\begin{landscape}
\begin{longtable}{|lccccccc|}
\caption[ ]{The list of effective temperatures from some 
authors (Morales et al. 2008, Jenkins et al. 2009, 
Wright et al. 2011, Stelzer et al. 2013, Lepine et al. 2013,
 Gaidos et al. 2014) for our sample of dM4 stars.} \\

\hline    &               &              &              &             &              &             &             \\
Star      &$T_{eff}$\footnotemark[1]&$T_{eff}$ & $T_{eff}$& $T_{eff}$ &  $T_{eff}$   & $T_{eff}$   & $T_{eff}$   \\
          &     (K)       &     (K)      &     (K)      &       (K)   &     (K)      &     (K)     &     (K)     \\
          & This Work     & Morales 2008 & Jenkins 2009 & Wright 2011 & Stelzer 2013 & Lepine 2013 & Gaidos 2014 \\
          &               &              &              &             &              &             &             \\
\endfirsthead
\caption{continued.}\\

\hline
          &               &              &              &             &              &             &             \\
Star      &$T_{eff}$\footnotemark[1]&$T_{eff}$ & $T_{eff}$& $T_{eff}$ &  $T_{eff}$   & $T_{eff}$   & $T_{eff}$   \\
          &     (K)       &     (K)      &     (K)      &       (K)   &     (K)      &     (K)     &     (K)     \\
          & This Work     & Morales 2008 & Jenkins 2009 & Wright 2011 & Stelzer 2013 & Lepine 2013 & Gaidos 2014 \\
          &               &              &              &             &              &             &             \\
\endhead  
\hline
\endfoot

GJ        &  3297         &       -      &      -       &      -      &       -      &     -       &     -       \\
GJ 1001A  &  3297         &       -      &      -       &      -      &       -      &     -       &   3351      \\
GJ 1005AB &  3315         &       -      &      -       &      -      &       -      &     -       &   3227      \\
GJ 1006A  &  3273         &       -      &      -       &      -      &       -      &    3250     &   3298      \\
GJ 1006B  &  3263         &       -      &      -       &      -      &       -      &    3100     &   3211      \\
GJ 1013   &  3304         &       -      &      -       &      -      &       -      &     -       &   3243      \\
GJ 1034   &  3250         &       -      &     3035     &      -      &       -      &     -       &     -       \\
GJ 1092   &  3251         &     3093     &      -       &      -      &       -      &     -       &     -       \\
\hline 
\end{longtable}
\end{landscape}
\normalsize
\footnotetext[1]{From the (R-I)$_C$ color.}

\end{document}